\documentclass[twocolumn,superscriptaddress,graphicx]{revtex4}
\usepackage{graphicx}

\begin{document}


\title{32 Bin Near-Infrared Time-Multiplexing Detector with Attojoule Single-Shot Energy Resolution} 



\author{Patrick Eraerds}
\email{Patrick.Eraerds@unige.ch}
\affiliation{Group of Applied Physics, University of Geneva, 1211 Geneva, Switzerland.}
\author{Enrico Pomarico}
\affiliation{Group of Applied Physics, University of Geneva, 1211 Geneva, Switzerland.}
\author{Jun Zhang}
\affiliation{Group of Applied Physics, University of Geneva, 1211 Geneva, Switzerland.}
\author{Bruno Sanguinetti}
\affiliation{Group of Applied Physics, University of Geneva, 1211 Geneva, Switzerland.}
\author{Rob Thew}
\affiliation{Group of Applied Physics, University of Geneva, 1211 Geneva, Switzerland.}
\author{Hugo Zbinden}
\affiliation{Group of Applied Physics, University of Geneva, 1211 Geneva, Switzerland.}


\date{\today}

\begin{abstract}
We present two implementations of photon counting time-multiplexing detectors  for near-infrared wavelengths, based on Peltier cooled InGaAs/InP avalanche photo diodes (APDs). A first implementation is motivated by practical considerations using only commercially available components. It features 16 bins, pulse repetition rates of up to 22 kHz and a large range of applicable pulse widths of up to 100 ns. A second implementation is based on rapid gating detectors, permitting deadtimes below 10 ns. This allows one to realize a high dynamic-range 32 bin detector, able to process pulse repetition rates of up to 6 MHz for pulse width of up to 200 ps. Analysis of the detector response at 16.5\% detection efficiency, reveals a single-shot energy resolution on the attojoule level.
\end{abstract}


\maketitle 

\section{Introduction}
Conventional semiconductor-based single photon detectors, i.e. single-pixel Geiger-mode avalanche photo diodes (APDs),  are used in very different domains where high sensitivity is demanded. Naturally they are binary detectors, giving a click or no-click in response to an incoming light pulse. While this behavior is desired in fields like quantum information or quantum cryptography,  it represents a restriction when light pulses containing a few to hundreds or thousands of photons have to be measured. The restricted dynamic requires the use of an attenuator and numerous measurement repetitions in order to infer the average pulse energy \cite{Paris}. Therefore a detector with the same sensitivity as single-pixel photon counting detectors, but with a larger response dynamic is desirable. It would strongly reduce the measurement time and allows one to infer single-shot estimates of the pulse energy. This can, in particular, be an advantage in biological imaging, where the number of applicable repetitions is restricted by the sensitivity of the specimen, or, in the characterization of mesoscopic quantum states \cite{Pavel}.

Detectors that are capable of estimating the number of photons in a pulse on a single-shot basis are commonly called "photon number resolving detectors". The resolution on that number strongly depends on the efficiency of the detector. Examples for the visible wavelength regime are the silicon photo multiplier (SiPM) \cite{Gomi, Eraerds}, time-multiplexing detector \cite{Fitch03,Achilles04} and visible light photon counter (VLPC) \cite{VLPC1, VLPC2} , and for the near-infrared the multi-pixel array \cite{Dauler07}, frequency upconversion combined with SiPM \cite{Pom}, avalanche height discrimination detector \cite{ShieldsPNR} and transition edge sensors (TES) \cite{Nam08,Fukuda09}. In the near-infrared, detectors either suffer from high dark count rates \cite{Dauler07}, low efficiency \cite{Pom}, poor dynamic \cite{ShieldsPNR} or depend on cryogenic cooling \cite{Nam08,Fukuda09}.

In this paper we present two implementations of time-multiplexing detectors for the near-infrared wavelength regime (1100 - 1650 nm), exhibiting low dark count, compact cooling, high dynamic and the capability of processing high pulse repetition rates. 

Time-multiplexing means that a pulse with a large number of photons is split into well separated weaker pulses by means of fiber delay loops (Fig.\ref{fig:setup}), subsequently detected by single-pixel InGaAs/InP APDs. This kind of detectors has been implemented so far only in the visible wavelength regime using Silicon APDs \cite{Fitch03,Achilles04}. Up to three fiber loops (22, 44 and 88 m), distributing the initial photons on 16 different bins, were demonstrated. Due to high fiber loss at visible wavelengths ($\geq$ 6 dB/km), these detectors are not efficiently scalable to a higher number of loops.

Our first detector version uses commercially available Geiger-mode APDs (ID200, IDQ) and represents an easy-to-implement and easy-to-control scheme. The second version applies rapid gating detectors \cite{Jun,JunSpie,ShieldsRapid,Inoue} with drastically decreased afterpulsing, allowing system scalability due to low fiber loss at near-infrared wavelengths ($< 0.2 $ dB/km).

The remainder of this paper is organized as follows: Section \ref{sec:setups} describes both setups and their characterization using coherent pulses. In Section \ref{sec:Resolution} we evaluate the energy resolution for single and multi-shot measurements and section \ref{sec:Concl} contains our conclusion. 

\section{Setup}
\label{sec:setups}
\begin{figure}[h]
\begin{center}
\includegraphics[width=8.5cm]{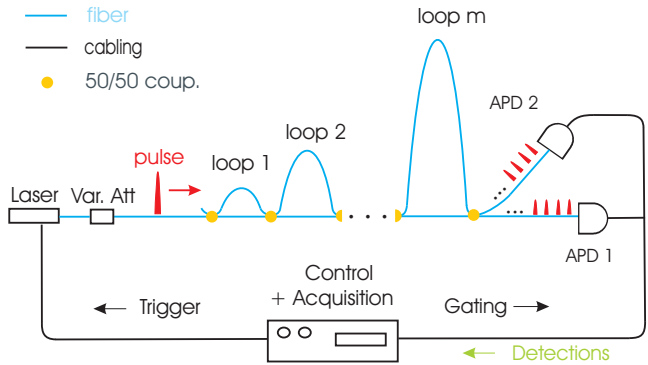}%
\caption{Schematic of the common setup, comprising a time-multiplexer (m fiber loops), two InGaAs/InP detectors and a central control and acquisition unit, which is responsible for accurate synchronization of laser pulses and detector gates as well as for the registration of detections.}
\label{fig:setup}
\end{center}
\end{figure}

The common setup for both multiplexing detectors is shown in Fig.\ref{fig:setup}. It consists of two InGaAs/InP APDs, a time-multiplexer constructed by single-mode fibers and a control and acquisition system. The latter is responsible for the synchronized timing of pulses and gates, and the acquisition of detections. The multiplexing part includes $m$ fiber loops (m=3 for the first and m=4 for the second implementation) fusion spliced via 50/50 couplers. The initial pulse is split into $2^{m+1}$ weaker pulses, subsequently called bins, $2^m$ directed to each detector. In the ideal case loop $i$ has at least twice the length of the previous one, guaranteeing well separated bins.
The absolute loop lengths, however, have to be chosen in accordance with the detector characteristics. Here, the tolerable afterpulsing is crucial. To keep its impact low, the distance between adjacent bins (deadtime) has to be chosen sufficiently large.  

\subsection{Conventional gating detectors}
\label{Conv}
\begin{figure}
\begin{center}
\includegraphics[width=8.5cm]{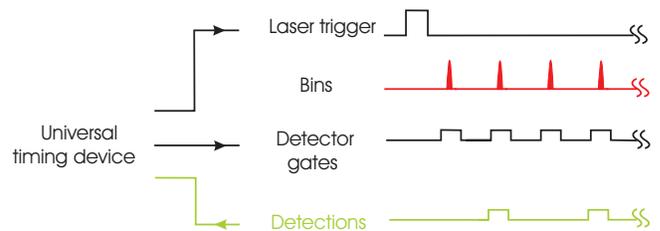}
\caption{Timing of conventional gating time-multiplexing detector. The universal timing device (UTD) sends laser trigger pulses with a fixed frequency, which generates 16 bins after the multiplexer. Gate signals are sent with an adequate delay in order to activate the detector at the pulse arrivals (bins). The number of occurred detections is then registered by the UTD.}
\label{fig:TimingConv}
\end{center}
\end{figure}

Two Peltier cooled InGaAs/InP  APDs (ID200,IDQ), using a conventional gating technique with applicable gate widths between 1 ns and 100 ns, are gated exclusively during the arrival of each bin, see Fig.\ref{fig:TimingConv}. This is facilitated by an universal timing device (UTD) (Chronologic Legato), which sends a trigger signal to the laser and with a certain delay a gate pattern to the APDs. The timing of the pattern can be defined by software in correspondence to the bins of the multiplexer. The timing accuracy of the module is typically 300\,ps. The UTD also acquires the number of detections by means of two additional ports, subsequently stored on the computer (USB connection).  

We choose a deadtime of 5 $\mu s$ at a detection efficiency of $\eta=$10\%. The afterpulse probability in a 20 ns gate after a detection gate is equal to 9\%. The dark count probability is $8\cdot10^{-6}$ per ns of gate. The deadtime setting demands a first loop of at least 1 km. In total we implement 3 loops of 1 km, 2 km and 5 km, yielding temporal delays of $5, 10$ and 25 $\mu$s and 16 bins. The overall transmission loss, averaged over all bins, is 1.44 dB.
 
We characterize the detector by sending coherent pulses of 1\,ns-width provided by a diode laser at 1559\,nm. They can be attenuated down to a well defined average number of photons per pulse by means of a calibrated variable attenuator.  The maximally processable laser pulse repetition rate is 22\,kHz, limited by the total length of the pulse train of 45 $\mu s$. 

\begin{figure}
\begin{center}
\includegraphics[width=6 cm]{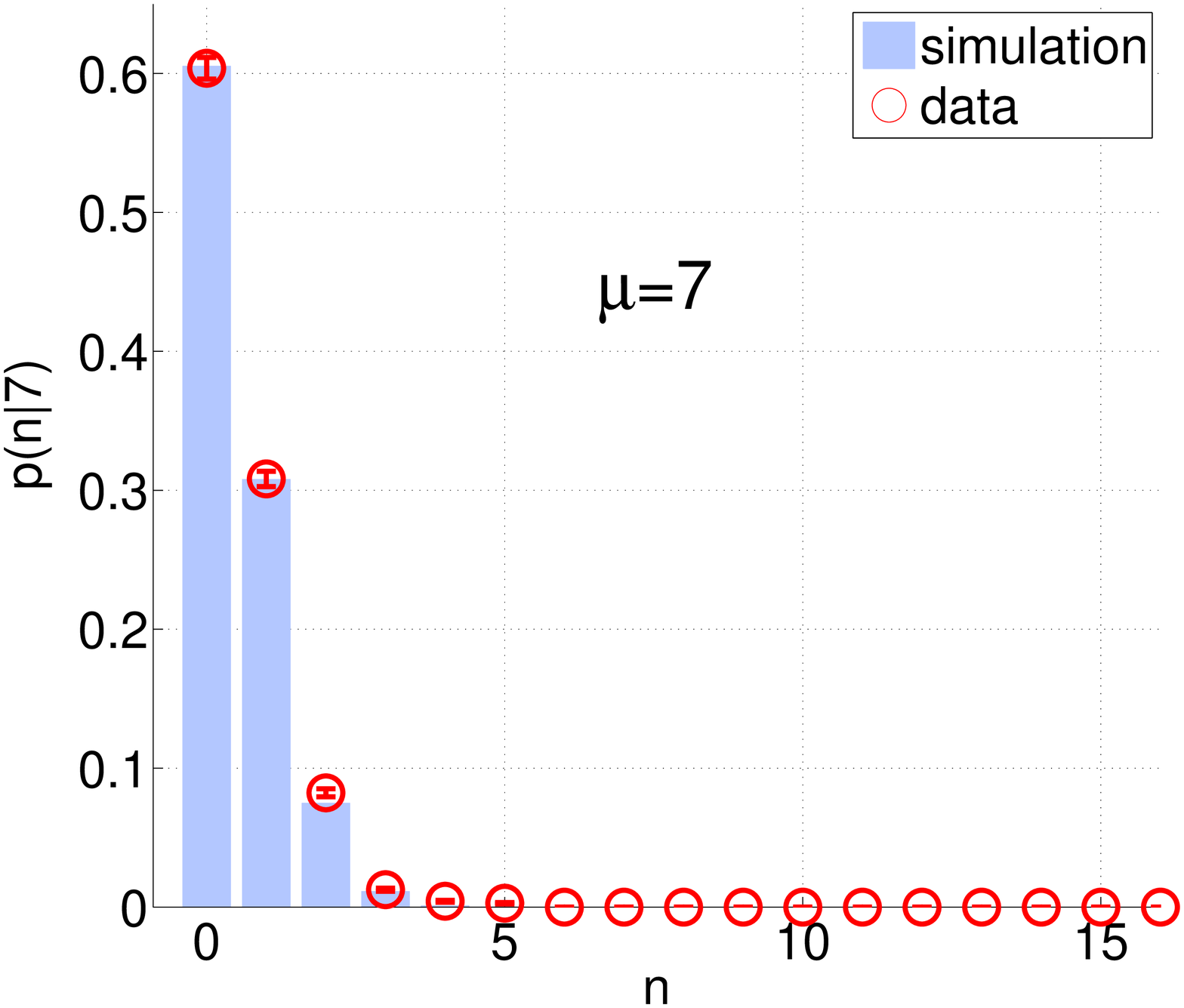}
\includegraphics[width=6 cm]{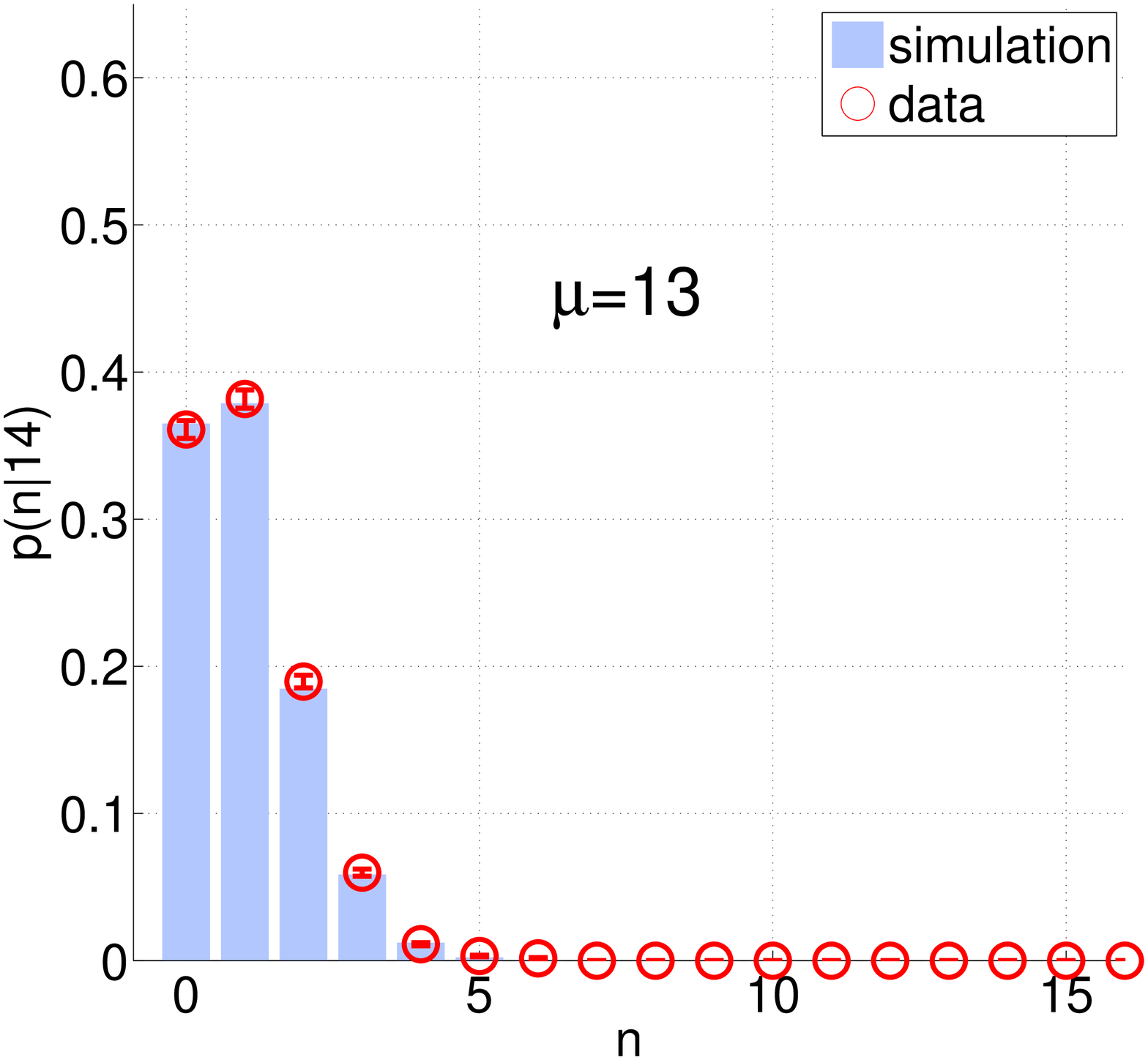}
\includegraphics[width=6 cm]{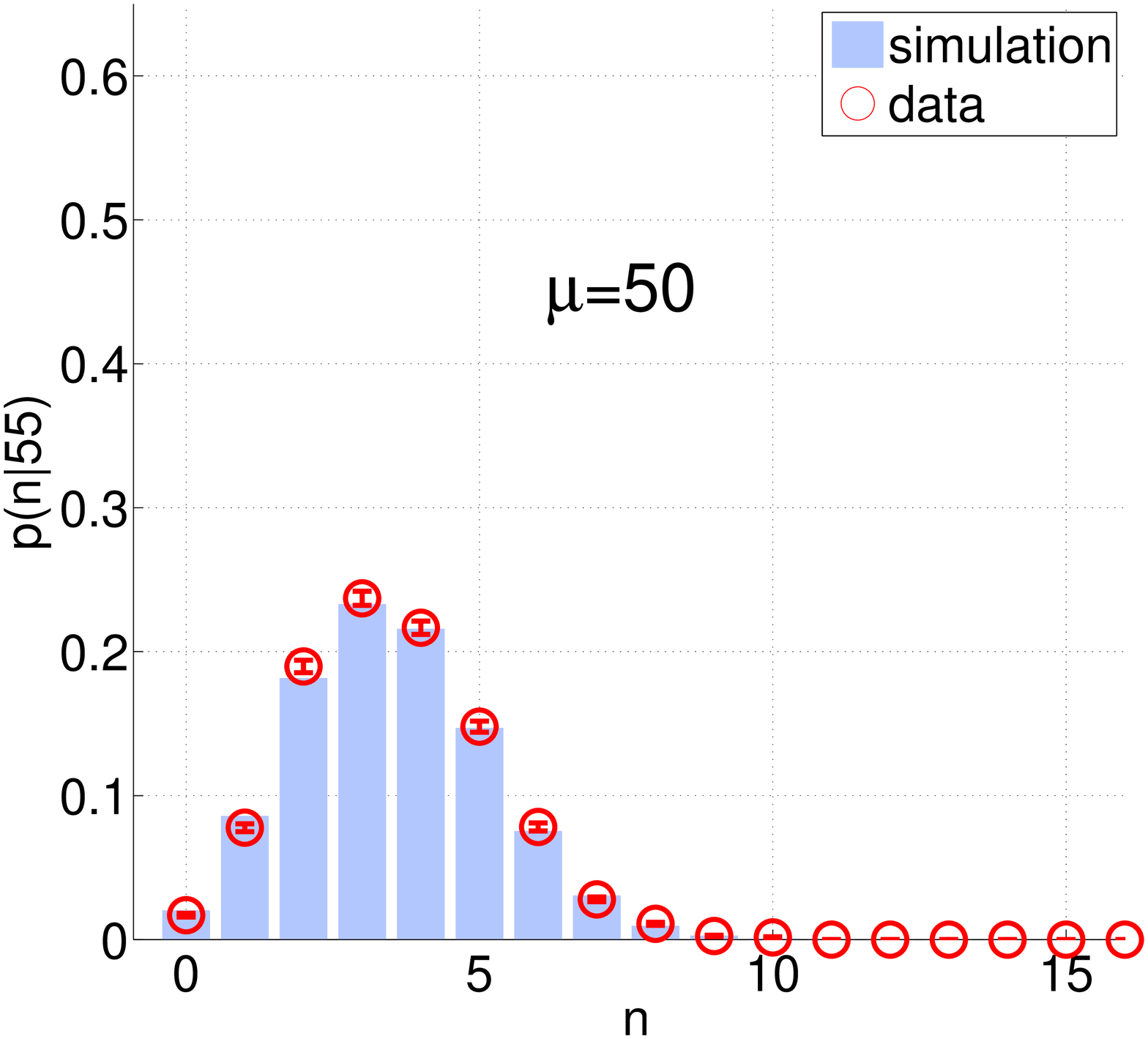}
\includegraphics[width=6 cm]{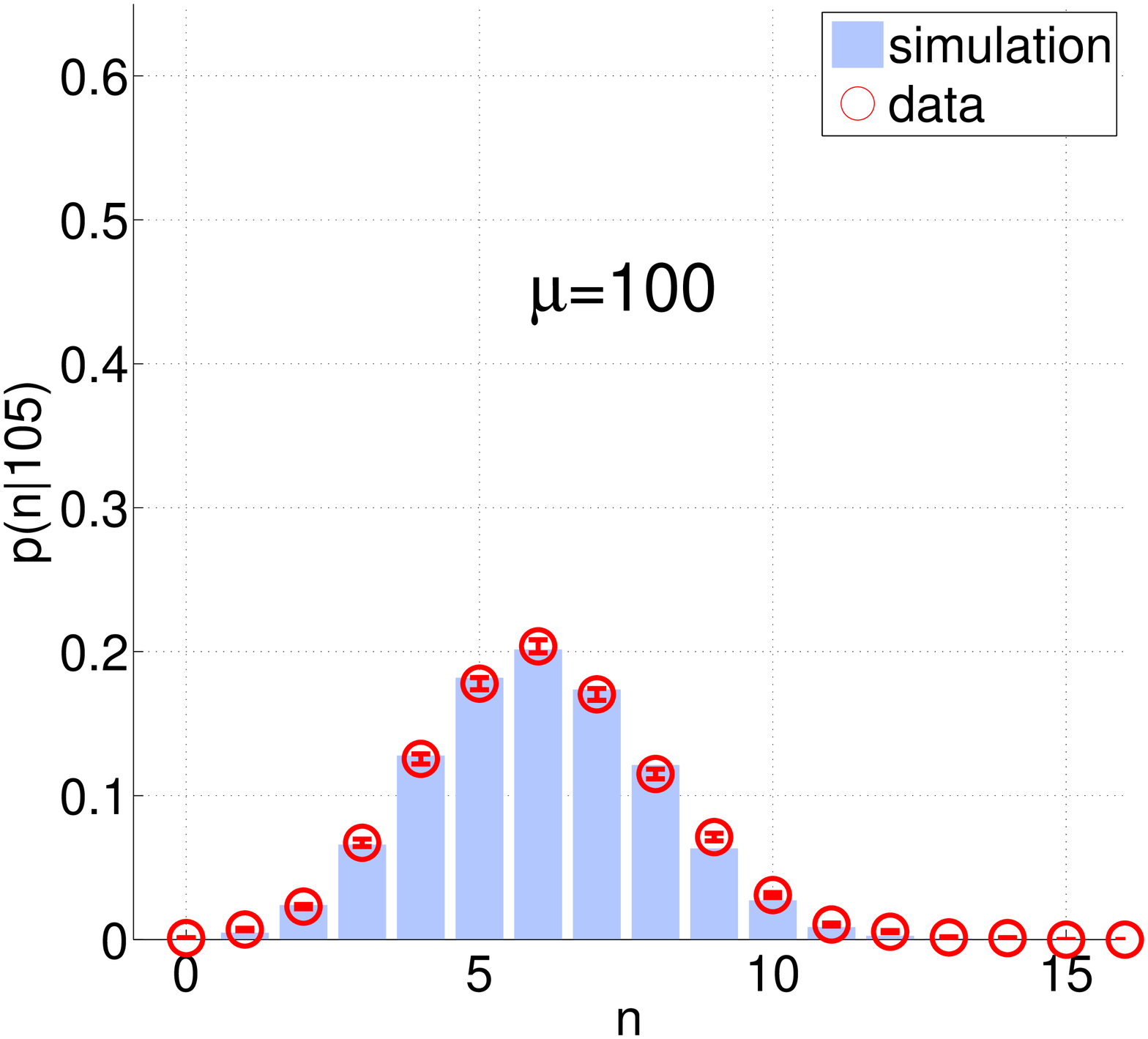}
\caption{Probability distributions $p(n|\mu)$ of the number of detections $n$, when coherent pulses of average photon number $\mu=$ 7, 14, 55 and 105 are sent onto our first time-multiplexing detector.}
\label{fig:resultsCon}
\end{center}
\end{figure}

We acquire data typically for a few seconds until sufficient statistics are obtained. The laser repetition rate in these measurements is 10\,kHz. Fig.\ref{fig:resultsCon} shows the probability distributions $p(n|\mu)$ (red circles with error bars) of the number of detections $n$ obtained by sending weak coherent pulses of average photon numbers $\mu=$ 7, 13, 50 and 100, using a gate width of 20 ns. In order to evaluate our results, we perform a simple Monte Carlo simulation, modelling the fiber couplers as perfect 50/50 beam splitters and accounting for the overall transmission loss in the multiplexer as well as dark counts and detector efficiency. The $\mu-$value that best fits the experimental results (plotted in Fig.\ref{fig:resultsCon}) is on average 8\% higher than in the pulse that is sent on the detector. This offset can be explained by the fact that afterpulsing is not accounted for in the simulation. We conclude that even without a specific model for afterpulsing the detector behavior can be simulated in good agreement with the actual detector response. 

A possible extension of this detector, realizing a forth loop, would require an additional fiber of 10 km length. This does not only increase the demand for space, but also significantly decreases the maximal pulse repetition rate as well as the transmission. In the next section we discuss the realization of an experiment using detectors with significantly lower afterpulsing permitting to use smaller deadtimes and hence making smaller loops and higher pulse repetition rates applicable .

\subsection{Rapid gating detectors}
\begin{figure}[b]
\begin{center}
\includegraphics[width=9cm]{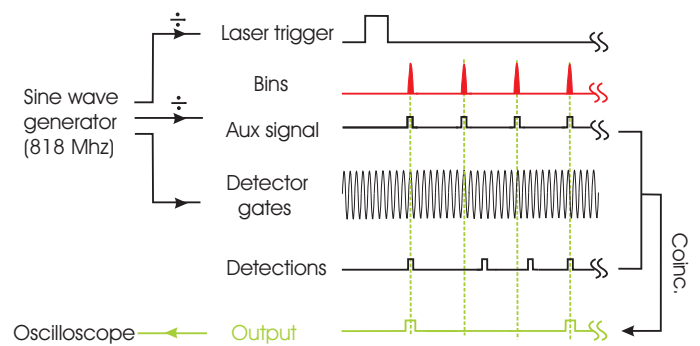} 
\caption{Timing schematic of rapid gating time-multiplexing detector. Central timing unit is a sine wave generator ($f=818$ MHz) that is responsible for detector gating as well as for laser triggering (with a divided frequency $(\div)$). An auxiliary signal at $f/8$ is used to perform coincidence counting, working as an additional filter against noise. The coincidence signals are then registered in an oscilloscope, which is triggered at the laser repetition frequency (connection not drawn).}
\label{fig:timingRap}
\end{center}
\end{figure}
Rapid gating detectors  \cite{Jun,JunSpie,Inoue} use very short gates in order to limit the total avalanche current. This drastically decreases afterpulsing and permits high gate rates. Practical deadtimes of about 10 ns allow to realize a time-multiplexer with fiber loops in the meter range. We implement 4 loops of  2m, 4m, 8m and 16 m generating  32 bins in total. The overall transmission loss, averaged over all bins, is 1.35 dB, mainly due to the large amount of splices. 

A sine wave generator ($f=818$ MHz, Agilent) is used as the central synchronization unit (Fig.\ref{fig:timingRap}). It applies gates to two Peltier cooled ($-40^\circ$C) JDSU InGaAs APDs (effective gate width of 200 ps) and triggers the diode laser (1550 nm, 30 ps pulses, PicoQuant) with a heavily divided frequency. Details about the gating technique  can be found in \cite{Jun,JunSpie}. In addition to these signals we generate an auxiliary signal at $f/8=102.25$ MHz whose period of 9.78 ns corresponds exactly to the distance of adjacent bins. It is synchronized with the arrival of the bins and used to perform coincidence counting. By this means, detections that occur between bin arrivals are filtered out. The total length of the bin train is 150 ns, which permits pulse repetition rates of up to 6 MHz. The number of coincidence detections for each laser pulse are registered on an oscilloscope (Lecroy 8600A). These numbers are recorded and analyzed later on.

The dark count probability per coincidence gate, at a detection efficiency of $\eta=$16.5\%, is equal to $1\cdot10^{-5}$ for the first and $5\cdot10^{-5}$ for the second detector. This efficiency value is close to the practical limit of our detectors and even a very small increase of the bias voltage leads to a significant growth of dark counts. With this setting we thus obtain a probability of $1\cdot10^{-3}$ for a dark count to appear in a single-shot measurement (32 bins), which is negligible.

The full characterization of afterpulsing is complicated \cite{Jun,JunSpie}. Here we perform a relatively simple measurement to roughly estimate the impact of afterpulsing. We use a single detector and send pulses with $\mu=$100 at a frequency of 512 kHz, which is well below the maximum of 6 MHz. Now we regard the number of detections in the 16 coincidence gates which are synchronized with the arriving bins and the number of detections in the immediately following 16 coincidence gates where no light is awaited. After 3000 pulses the sum of detections that occurred in the first 16 gates is 12806, while it is equal to 144 in the second 16 gates.
This would thus add 1\% of detections to the succeeding 16 coincidence gates in the limiting case of 6 MHz repetition rate, when signal photons are possibly present in each coincidence gate. For most applications this contribution is insignificant.

\begin{figure}
\begin{center}
\includegraphics[width=6.5 cm]{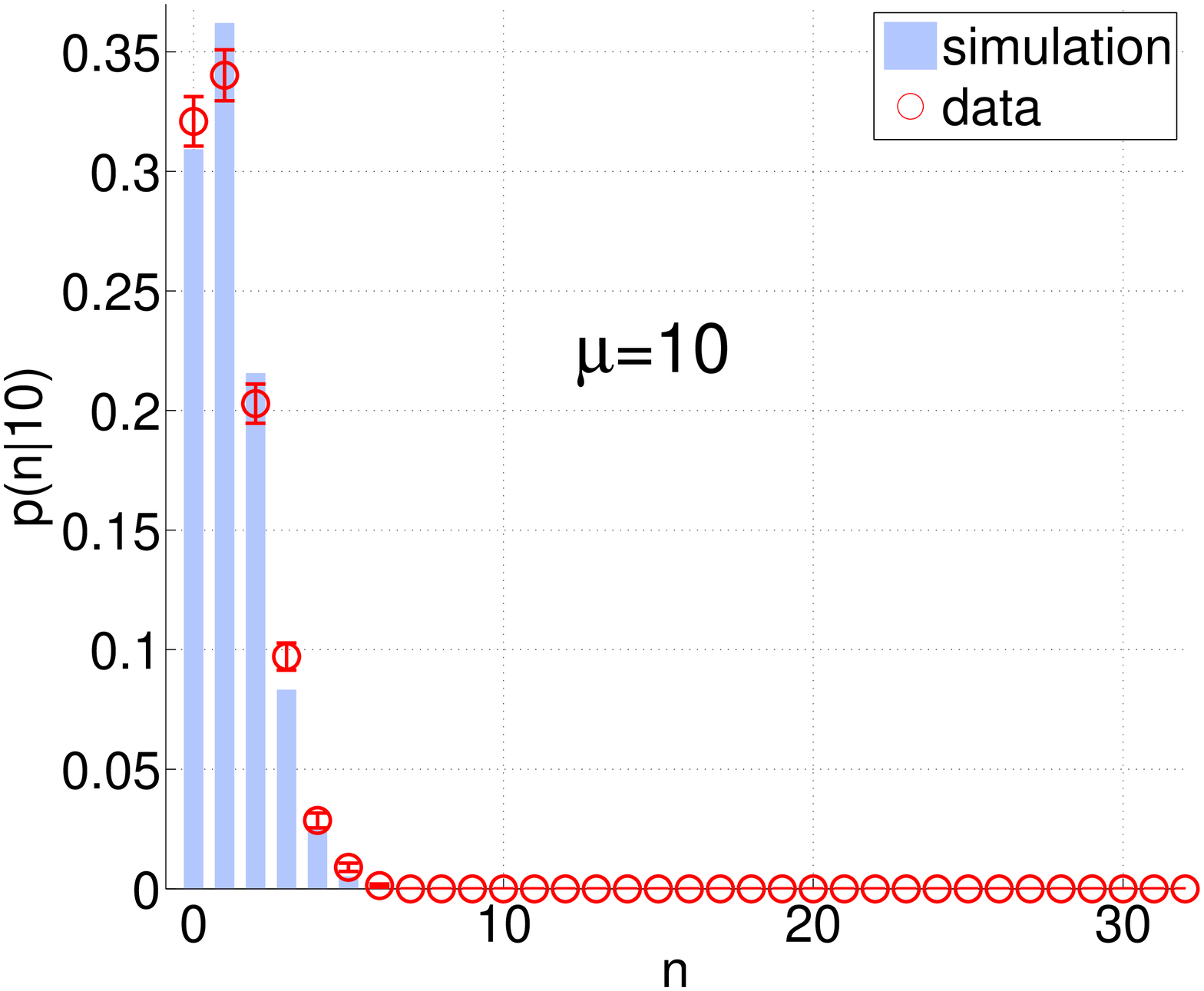}
\includegraphics[width=6.5 cm]{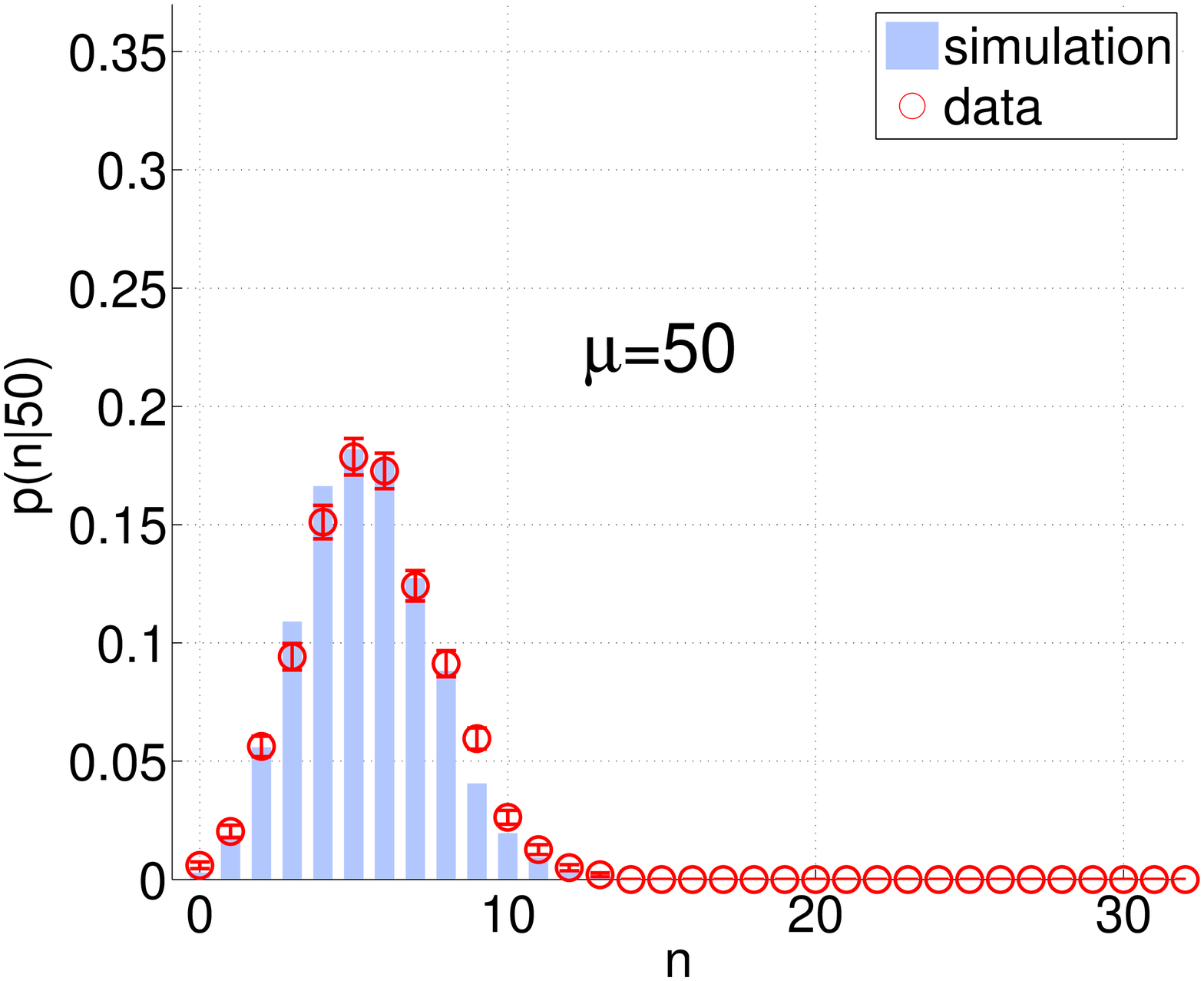}
\includegraphics[width=6.5 cm]{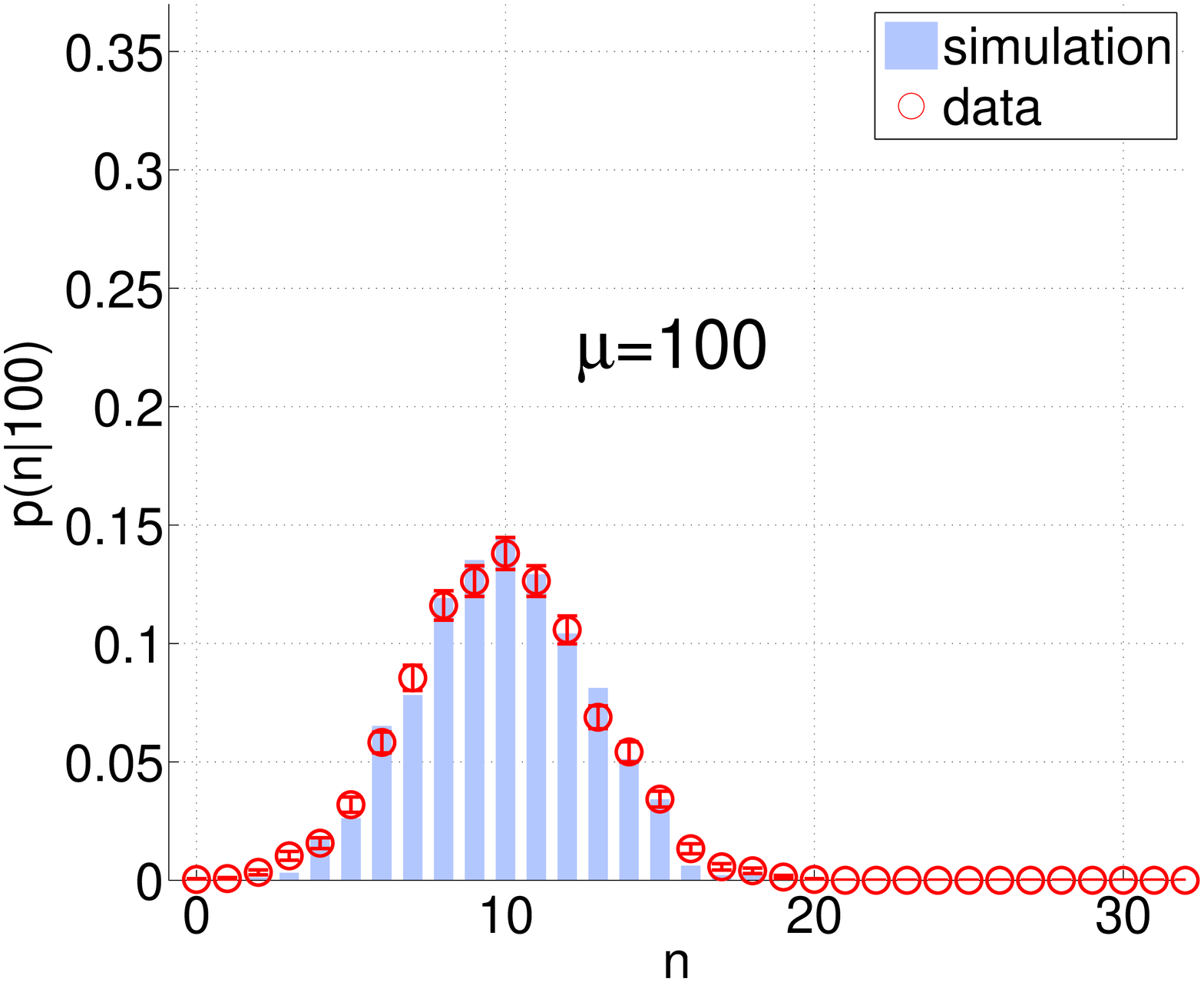}
\includegraphics[width=6.5 cm]{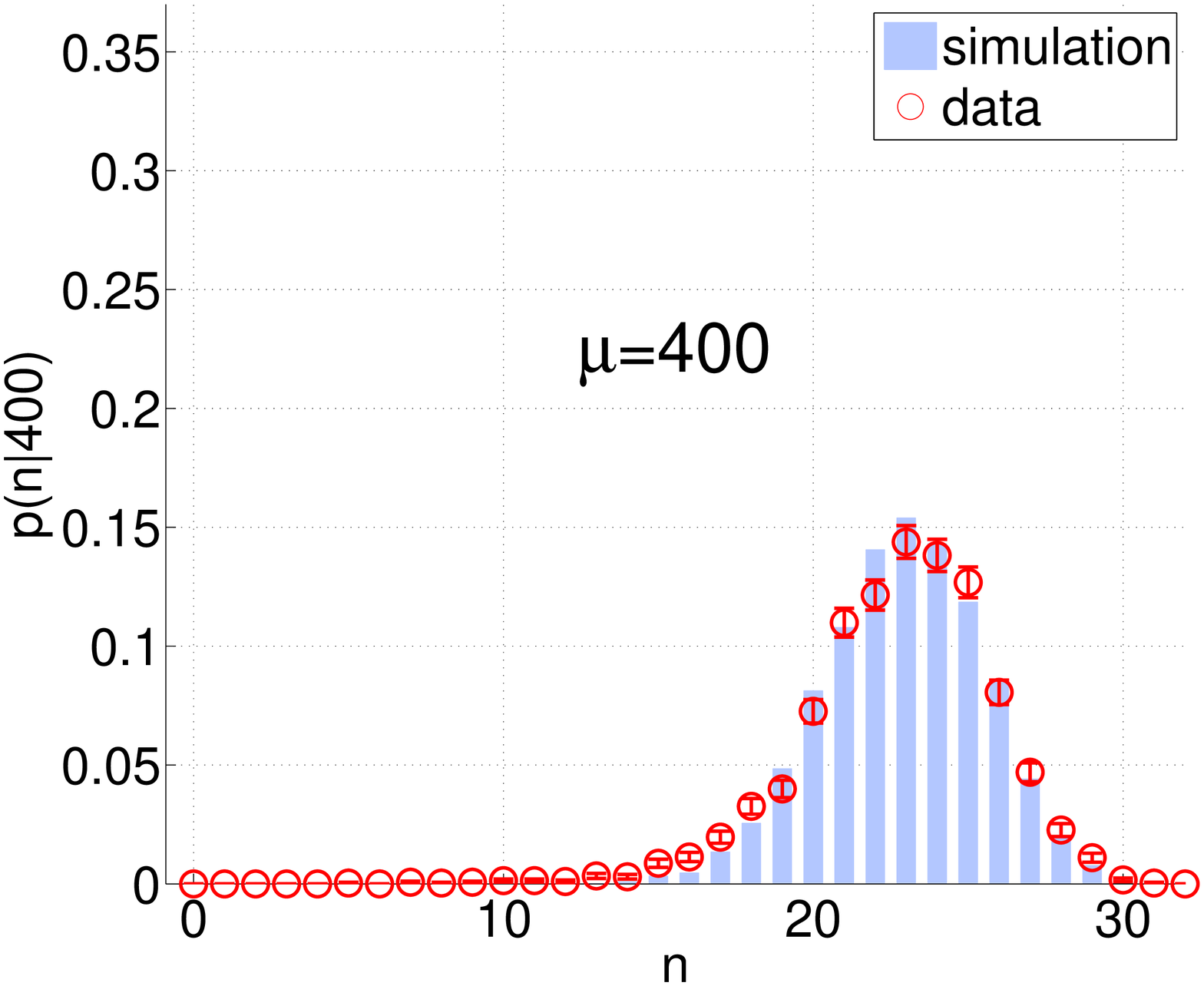}
\caption{Probability distributions $p(n|\mu)$ of the number of detections $n$, when coherent pulses of $\mu=$ 10, 50, 100 and 400 are sent on the rapid gating multiplexing detector.}
\label{fig:resultsRap}
\end{center}
\end{figure}

As done in the previous implementation, we characterize the rapid gating time-multiplexing detector by sending weak coherent pulses with different mean number of photons $\mu$. The acquisition using the oscilloscope takes a few minutes in order to obtain sufficient statistics. Results for $\mu=$10, 50, 100 and 400 are shown in Fig.\ref{fig:resultsRap}. The Monte Carlo simulation now includes an additional adaptation of the detection efficiency at high $\mu$-values. An undershoot effect after a strong avalanche, caused by more than one photon, can lead to a non-discrimination of an avalanche in the succeeding bin. To obtain the optimal fitting result the single photon detection efficiency has to be adapted from 16.5 \% for low, to 14.5 \% for high $\mu$-values. Afterpulsing is not accounted for and no offset like in the conventional gating case (Sec.\ref{Conv}) is observed. With this adaptation, measurement results and simulation are in quite good agreement, showing in particular, that afterpulsing does not play an important role.

\section{Energy resolution}
\label{sec:Resolution}
In the following discussion we will concentrate on the rapid gating implementation due to its higher efficiency and dynamic range. The same analysis can of course be performed in the same manner for the other implementation.

The complete set of conditional probabilities $p(n|\mu)$ would perfectly characterize our detector. However, even the measurement of closely spaced discretized $\mu \in \left\{1, 2, 3, ..., \mu_{max}\right\}$ would be very time consuming 
and not practical.
Since our simulation agrees very well with the experimental data, we fill the detector response matrix $M$, where an element $M_{\mu,n}:=p(n|\mu), \mu \in \left\{1, 2, 3, ..., \mu_{max}\right\}$, with calculated values between the measured supporting points $\mu=$10, 50, 100, 200 and 400.
The upper plot of Fig.\ref{fig:SingleSh} shows a graphical representation of $M$ up to $\mu=150$.

\subsection{Single-shot resolution}
Graphically it is quite evident that one can read $M$ as well the other way round, meaning that we assume a certain number of detections $n$ and want to infer an estimation of the average number of photons $\mu$ of the initial pulse, i.e. the conditional probabilities $p(\mu|n)$ for a given $n$. 
\begin{figure}[]
\begin{center}
\includegraphics[width=9 cm]{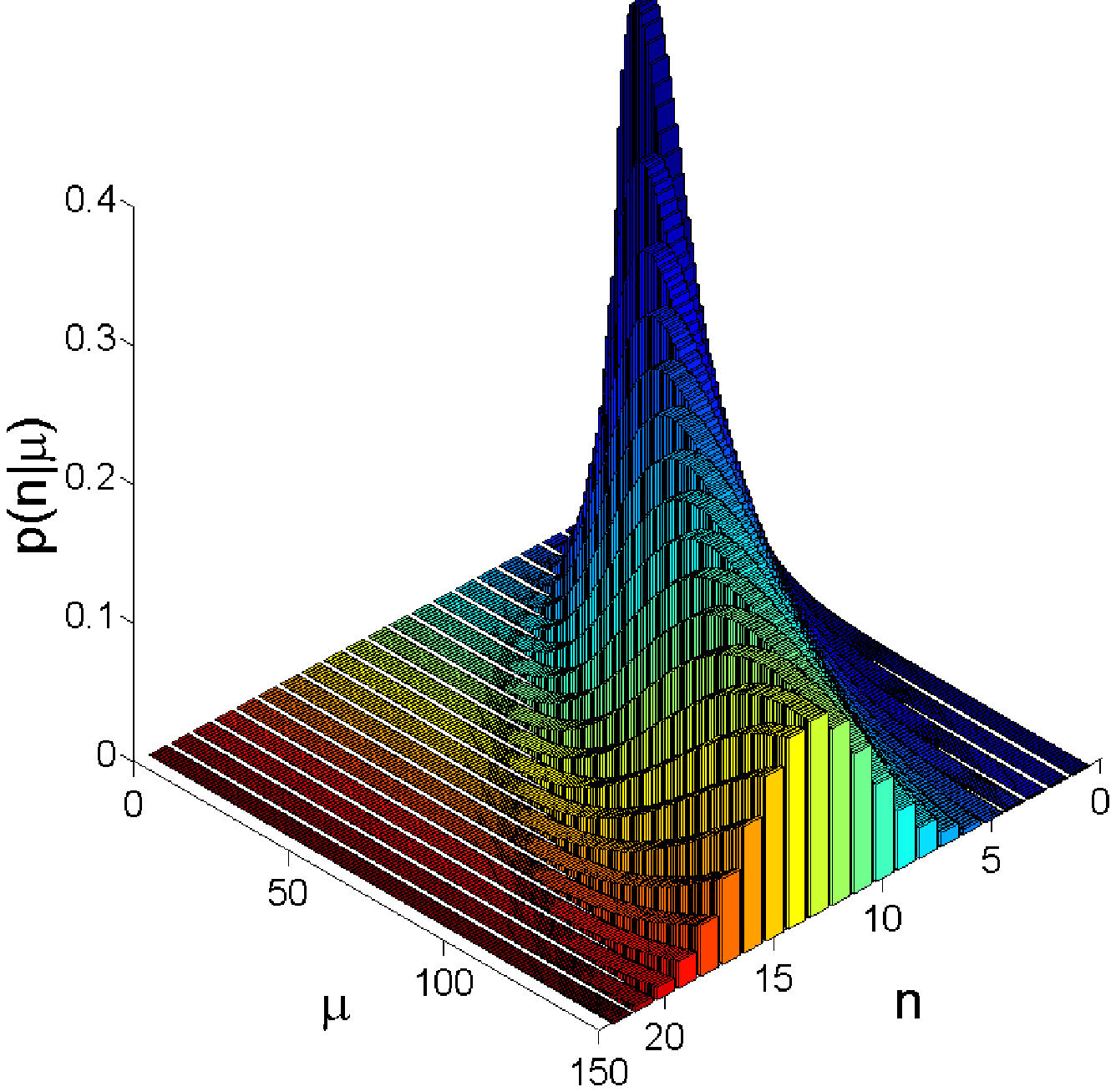}
\includegraphics[width=8 cm]{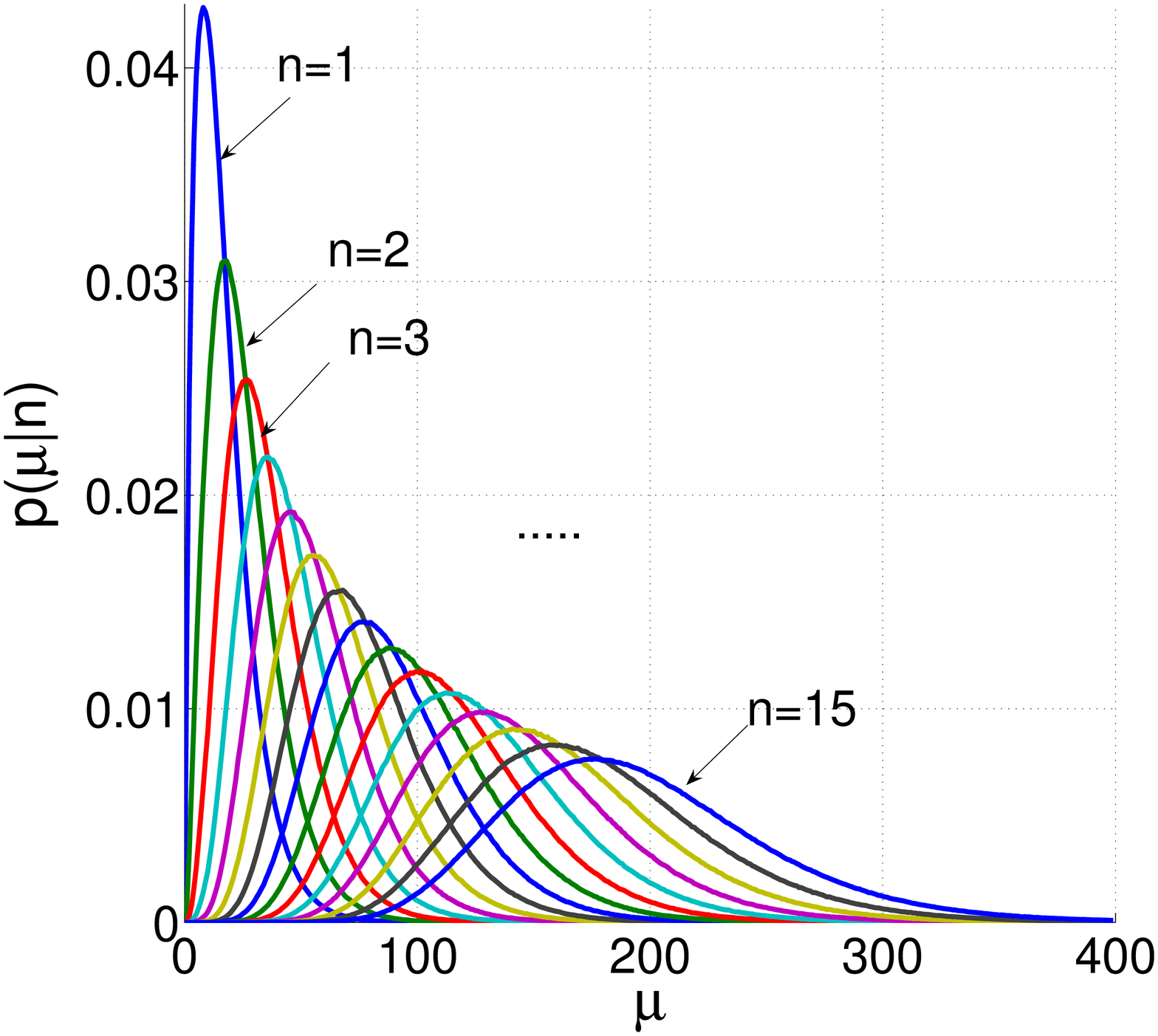}
\caption{Top : Graphical representation of detector response matrix $M$ for coherent pulses from $\mu$=0 to 150. Bottom : Distribution of $\mu$-values for single-shot detection events from 1 to 15 detections.}
\label{fig:SingleSh}
\end{center}
\end{figure}

Formally we infer the $p(\mu|n)$ by using the Bayesian theorem and assuming a uniform distribution for $p(\mu)$ in the range of characterization ($\mu=0,1,...,\mu_{max}$):

$$p(\mu|n)=\frac{p(\mu\& n)}{p(n)}=\frac{p(n|\mu)\cdot p(\mu)}{p(n)} =\frac{p(n|\mu)\cdot p(\mu)}{\sum_{\mu'=0}^{\mu_{max}}p(\mu')\cdot p(n|\mu')}$$
\begin{equation}
\label{equ1}
=\frac{p(n|\mu)}{\sum_{\mu'=0}^{\mu_{max}} p(n|\mu')}.
\end{equation}
In the last step the assumption of uniformly distributed $\mu$ is used. The lower plot of Fig.\ref{fig:SingleSh} shows these distributions for different $n$ from 1 to 15.

The choice of $\mu_{max}$ ($\mu_{max}=400$ in our case) is of course arbritrary, but it is important to ensure that the distributions  $p(\mu|n)$ in Eq.\ref{equ1} are independent of $\mu_{max}$. Consider for example the distribution $p(\mu|1)$ (bottom plot of Fig.\ref{fig:SingleSh}): the total probability for a $\mu>100$ to produce 1 detection is so small that plotting $p(\mu|1)$ for $\mu_{max}=100$ and $\mu_{max}=400$ leads to the same result, thus is stable for a $\mu_{max}>100$. In our case ($\mu_{max}=400$) the same argument holds for up to 15 detections. Single-shot events producing more than 15 detections need to be discarded, since the contributions from $\mu$-values larger than 400 is not negligible. If one wants to use this detector for pulses with higher average photon number, $\mu_{max}$ has to be adapted accordingly.
We note that if respecting this rejecting rule, the assumption of a uniform distribution of $\mu$-values is the most conservative (= no information about incoming pulse, except $\mu\leq \mu_{max}$).

Fig.\ref{fig:SingleSh} (bottom) reveals that even with a single pulse measurement (number of detections between 1 and 15), the range of possible average photon numbers $\mu$ is quite restricted. For example in case of only one detection ($p(\mu|1)$), one can state on a 90 \% confidence level, that the initial pulse had a $\mu$ of 8 $\pm^{28}_5$, corresponding to an energy resolution of 4.2 attojoule (equivalent to 33 photons at 1550 nm).

High resolution classical Joulemeters exhibit energy resolutions of down to 10 femtojoule, but are limited by pulse repetition rates of a few kHz \cite{spectrumdet}. High speed Joulemeters can process rates up to 100 kHz, but have energy resolutions on the nanojoule level. Our detector combines at the same time, high repetition rate of up to 6 MHz and high single-shot energy resolution on the attojoule level. We note that due to the short loop lengths of the rapid gating multiplexer ($\leq$ 30 m), the mode dispersion in a 50 $\mu m$ core (NA=0.2), graded index multimode fiber amounts to less than 10 ps and can therefore be used to increase the amount of captured light.

\subsection{Multi-shot resolution}
\begin{figure}
\begin{center}
\includegraphics[width=8 cm]{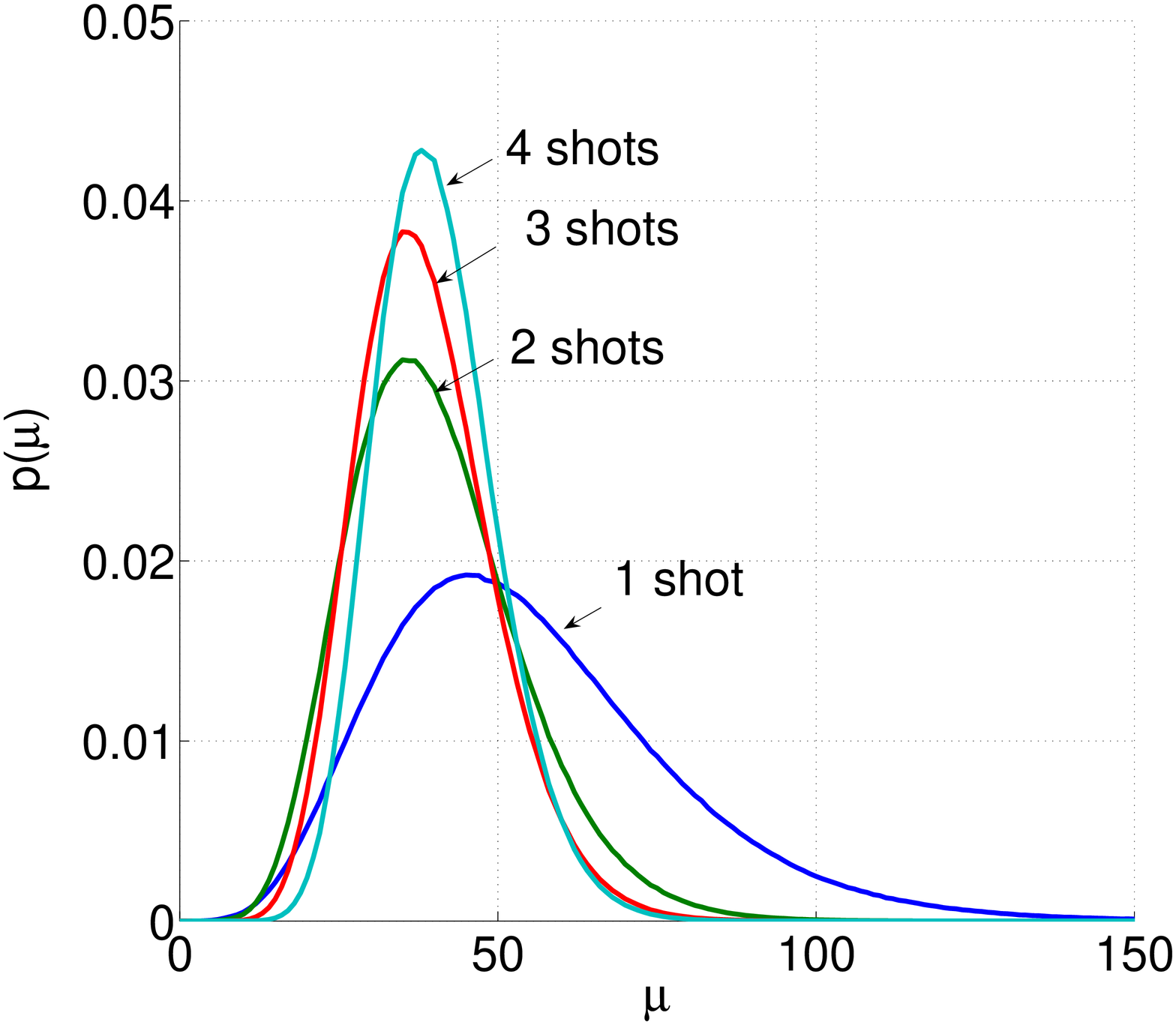}
\includegraphics[width=8 cm]{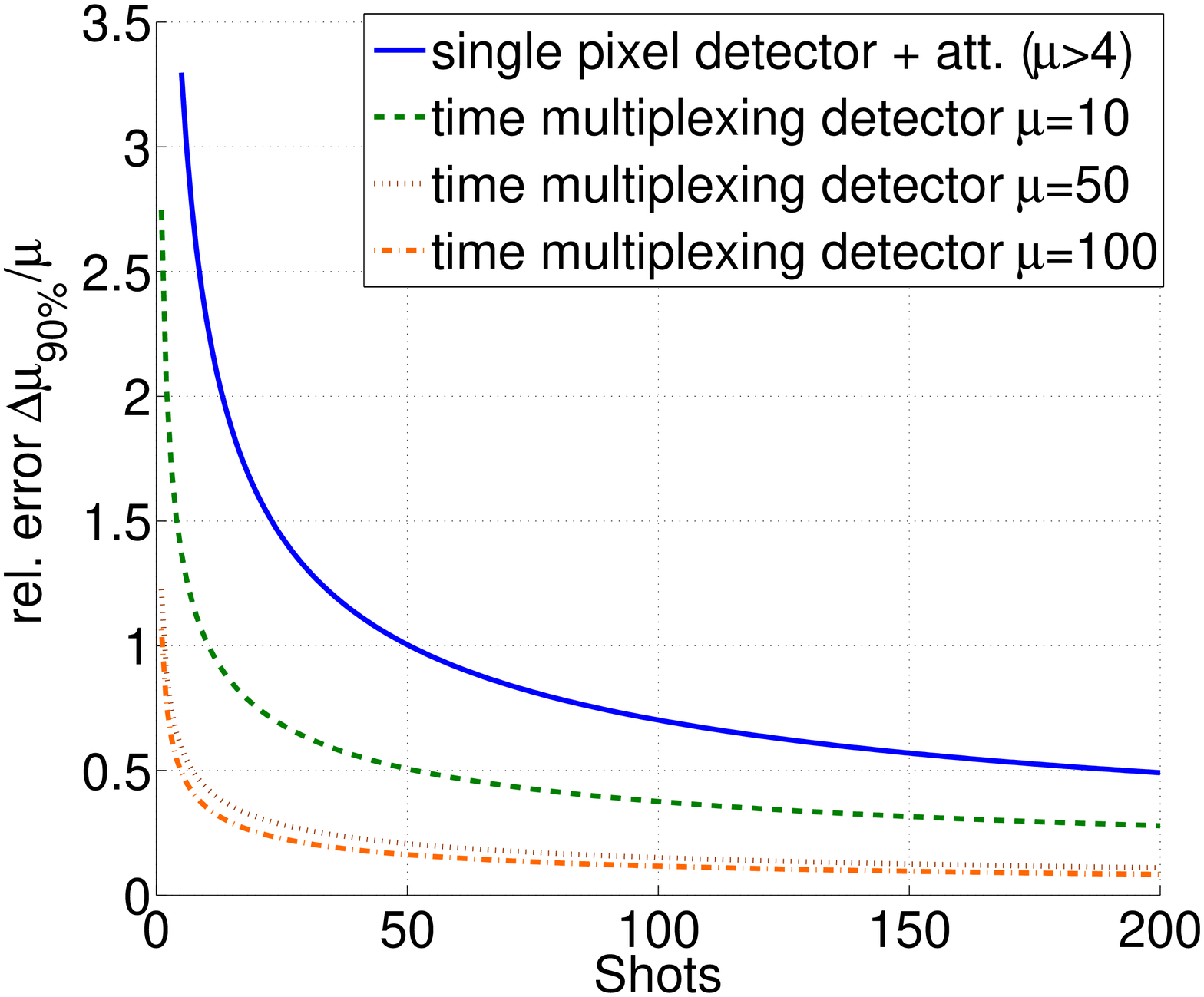}
\caption{Top: Evolution of the most probable $\mu$-values after a series of detection, here 5, 3, 4, 5. Bottom: Evolution of the relative error (assuming a 90 \% width) with number of shots, when subjected to repetitive coherent pulses of $\mu=10, 50$ and 100. As a comparison, we plot the relative error obtained by a single-pixel detector with variable attenuator to adjust a detection probability per gate of 50 \% (see text for details).}
\label{fig:multishots}
\end{center}
\end{figure}

Under the same assumptions as used before, we can obtain multi-shot distributions $p(\mu|n_1,n_2,...,n_k)$, where $n_1,n_2,...,n_k$ is a series of number of detections generated by coherent pulses with the same $\mu$-value.  Using the single-shot distributions $p(\mu|n)$ we calculate :
\begin{equation}
	p(\mu| n_1,...n_k)=c\cdot\prod_{i=1}^k p(\mu|n_i)
\end{equation}
where $c$ is a normalization factor ensuring that $\sum_{\mu=0}^{\mu_{max}}p(\mu| n_1,...n_k) =1$.
The additional information from each shot drastically narrows down the $\mu$-values that come into consideration. In Fig.\ref{fig:multishots} (top) we plot the evolution of the most probable $\mu$-values when a series of 5, 3, 4 and 5 detections is obtained. 
In the bottom plot the evolution of the relative error $\Delta\mu_{90\%}/\mu$ (90\% confidence level), as a function of the number of shots for different coherent pulses ($\mu = 10, 50, 100$), is shown. 

In addition to that, we plot the evolution of the relative error achieved by a gated single-pixel detector (without multiplexer) with the same detection efficiency ($\eta= 16.5\%$). 
The relation between the $\mu$-value of the incident pulse and the detection probability per gate $p_{det,gate}$ is given by
\begin{equation}
\label{SingleP}
	\mu=\frac{-1}{\eta\cdot\alpha} \mbox{ln}(1-p_{det,gate})
\end{equation}
The parameter $\alpha$ accounts for the case when an additional attenuator is used to avoid detector saturation.
$p_{det,gate}$ can be dynamically estimated via the ratio $N_{det}/N_{gate}$ of number of detections $N_{det}$ and the number of applied gates $N_{gate}$. Our results confirm the intuition that the relative error of $\mu$ decreases fastest for $p_{det,gate}$ about 50\%. This could be achieved experimentally for pulses with $\mu>4$ using a variable attenuator. We neglect cases with $\mu\leq 4$ in our discussion. The error on $\mu$, i.e. $\Delta \mu_{90\%}$, is propagated from a 90\% Poissonian error on $N_{det}$. Due to the constant $p_{det,gate}$ the relative error is independent of the strength of the incident pulses.
The contrary happens in the time-multiplexing setup. The higher the $\mu-$value, the faster the relative error decreases. For example, at $\mu=100$ a relative error of 0.1 is obtained after approximately 150 shots. The single-pixel detector needs roughly 4500 shots (not plotted) to achieve the same result. This implies a reduction of the number of shots by a factor of 30.  

\section{Conclusion}
\label{sec:Concl}
We find that Peltier cooled InGaAs/InP APD, sensitive in the near-infrared wavelength regime (1100-1650 nm), are suitable for the implementation of time-multiplexing detectors. We realize two different versions. The first one is an easy-to-operate and easy-to-control version based on commercially available detectors. It has a large flexibility in choosing gate widths, which in return permits to process a large range of laser pulse widths (up to 100 ns). Due to afterpulsing one has to respect relatively long delays between adjacent bins which necessitates the use of rather long fiber loops (order of kilometers). The maximal pulse repetition rate is 22 kHz. The second implementation is based on rapid gating detectors, exhibiting very low afterpulsing. Deadtimes of 10 ns allow the use of short fibers (a few meters) to realize the bin delays. 4 loops are used to create 32 bins, yielding a large response dynamic. The maximally attainable pulse repetition rate is 6 MHz and the effective gate width is fixed at 200 ps.
Analysis of the detector response to coherent pulses makes it possible to calculate single-shot and multi-shot energy resolutions of pulses of unknown energy. We find that in single-shot events the resolution can be as small as 4.2 attojoule.


%
%

%

\begin{acknowledgments}
 This work is supported by the Swiss NCCR "Quantum Photonics".
\end{acknowledgments}


\end{document}